\documentclass[%
 reprint,
 superscriptaddress,
 amsmath,amssymb,
 aps,
]{revtex4-1}

\usepackage{graphicx}
\usepackage{dcolumn}
\usepackage{bm}
\usepackage{xcolor}
\usepackage{float}
\usepackage{txfonts}

\begin{document}

\title{Hybrid neural network method of a multilayer perceptron
and autoencoder for $\alpha$-particle preformation factor 
in $\alpha$-decay theory}

\author{Jiaqi Luo}
\affiliation{Frontiers Science Center for Rare Isotopes, Lanzhou University, 730000 Lanzhou, People's Republic of China}
\affiliation{School of Information Science and Engineering, Lanzhou University, 730000 Lanzhou, People's Republic of China}
\author{Yang Xu}
\affiliation{School of Nuclear Science and Technology, Lanzhou University, 730000 Lanzhou, People's Republic of China}
\author{Xiaolong Li}
\affiliation{School of Nuclear Science and Technology, Lanzhou University, 730000 Lanzhou, People's Republic of China}
\author{Junxiang Wang}
\affiliation{School of Nuclear Science and Technology, Lanzhou University, 730000 Lanzhou, People's Republic of China}
\author{Yangjie Zhang}
\affiliation{School of Nuclear Science and Technology, Lanzhou University, 730000 Lanzhou, People's Republic of China}
\author{Jungang Deng}\email[Contact author: ]{dengjungang@ctgu.edu.cn}
\affiliation{
 Center for Astronomy and Space Sciences, China Three Gorges University, Yichang 443002, People's Republic of China}
\author{Fang Zhang}\email[Contact author: ]{zhangfang@lzu.edu.cn}
\affiliation{School of Nuclear Science and Technology, Lanzhou University, 730000 Lanzhou, People's Republic of China}
\author{Nana Ma}\email[Contact author: ]{mann@lzu.edu.cn}
\affiliation{Frontiers Science Center for Rare Isotopes, Lanzhou University, 730000 Lanzhou, People's Republic of China}
\affiliation{School of Nuclear Science and Technology, Lanzhou University, 730000 Lanzhou, People's Republic of China}

\begin{abstract}

The preformation factor quantifies the probability of $\alpha$ particles preforming on the surface of the parent nucleus in decay theory and is closely related to the study of $\alpha$ clustering structure. In this work, a multilayer perceptron and autoencoder (MLP + AE) hybrid neural network method is introduced to extract preformation factors within the generalized liquid drop model and experimental data. A $\mathcal{K}$-fold cross validation method is also adopted. The accuracy of the preformation factor calculated by this improved neural network is comparable to the results of the empirical formula.
MLP + AE can effectively capture the linear relationship between the logarithm of the preformation factor ($\text{log}_{10}P_\alpha^\text{exp}$) and the square root of the ratio of the decay energy ($Q_\alpha^{-1/2}$), further verifying that Geiger-Nuttall law can deal with preformation factor. 
The extracted preformation probability of isotope and isotone chains show different trends near the magic number, and in addition, an odd-even staggering effect appears. This means that the preformation factors are affected by closed shells and unpaired nucleons. Therefore the preformation factors can provide nuclear structure information. Furthermore, for 41 new nuclides, the half-lives introduced with the preformation factors reproduce the experimental values as expected. Finally, the preformation factors and $\alpha$-decay half-lives of $Z = $ 119 and 120 superheavy nuclei are predicted.

\end{abstract}

\pacs{23.60.+e, 23.70.+j, 07.05.Mh}

\maketitle


\section{Introduction}
\label{Introduction}
The study of superheavy elements and long-lived superheavy nuclei is an essential content for exploring the limits of the existence of electric charge. An important frontier in current nuclear physics is the synthesis of superheavy elements with proton numbers $Z >$ 118 \cite{WOS:000413521100001,WOS:000369773700001}. How to determine the identity of newly synthesized elements is particularly important. Currently, it is generally believed that $\alpha$ decay is almost the only means to identify the synthesis of new nuclides \cite{PhysRevC.110.044302,plus+1}. Therefore, a reliable $\alpha$-decay theory is particularly important. The development of $\alpha$-decay theory focuses mainly on barrier curves, decay energy \cite{WOS:000258423200009,WOS:000310091000002}, the nuclear medium effect \cite{RenPLB}, the nuclear deformation \cite{ GuoNPA} and $\alpha$-particle preformation factor \cite{ WOS:000393117100025,ZhangPRC}, and so on. Many empirical formulas \cite{WOS:000306845800005,QiChongPRL2009,WOS:000925464300007,PhysRevC.101.034307}, the macro-micro and microscopic models \cite{WOS:000258423200009,Buck,Xu,Moller,Sobiczewski,Ren,Royer,Poggenburg,Zhang1,Zhang2} have emerged in theoretical research on $\alpha$ decay. 

The success of machine learning in artificial intelligence \cite{CarleoRMP} has led to a new research paradigm for nuclear physics. Interdisciplinary research between nuclear physics and machine learning (ML) covers many frontiers of nuclear physics, such as atomic nucleus mass \cite{Niu, MaCPC}, nuclear reactions \cite{WOS:000744189400005,li2024machine}, equation of the state of nuclear matter \cite{WOS:001016207900001,WOS:000500331400002}, high-energy heavy-ion collisions \cite{Pang1}, nuclear deformation
and spectroscopic properties \cite{Lasseri}, $\alpha$-decay \cite{WOS:000925464300007,plus+2,plus+3, WOS:001304476900001,WOS:000611523900001}, $\beta$ decay \cite{WOS:000271352900045,WOS:000470857400001} and so on. Focusing on the interdisciplinary of ML and $\alpha$-decay theory, there are many common machine learning (ML) methods in nuclear physics research, including the radial basis function neural network (RBFN) \cite{WOS:000925464300007}, 
the multilayer perceptron (MLP) neural network \cite{WOS:001304476900001}, the Bayesian neural network (BNN) \cite{WOS:000470857400001} and the decision trees (DTs) \cite{WOS:000705844800002}, and so on. These current studies show that the advantages of ML are generally demonstrated in improving the accuracy of theory and applying it to the extraction of physical information. In addition, it allows for appropriate extrapolation and prediction of some physical quantities. 

At present, cross-disciplinary research between the two disciplines mainly faces breakthroughs in three aspects: (i) developing physics-driven machine learning methods to increase the interpretability of machine learning methods for physical research; (ii) mining multiple types of machine learning methods and verifying their applicability in the field of nuclear physics; (iii) exploring the applicability of machine learning in as many key problems in nuclear physics as possible and strengthening the possibility of new paradigms in machine learning research.

On the topic of $\alpha$-decay theory, $\alpha$-particle preformation factor represents the probability of an $\alpha$-cluster formation on the surface of the decaying parent nucleus. However, this factor is treated differently in various decay theories. In the cluster model, the $\alpha$ formation amplitude is usually treated as a constant, which improves the accuracy of the description of the half-life to some extent, but it is weak in capturing the microscopic information of nuclear structure \cite{WOS:A1993LE34800002,Buck,WOS:000239511600012}. Consequently, significant systematic deviations can arise between calculated and experimental $\alpha$-decay half-lives for the nuclei around the shell. Some recent work by Zhang and Deng \cite{WOS:000647421500074} has made good progress in extracting the preformation factors using a semi-empirical formulas, offering improved alignment with observed decay properties. Currently, one of the important divergences between theoretical $\alpha$-decay half-lifes and experimental data is how to calculate the $\alpha$-particle preformation factor. Machine learning has advantages in handling the differences between two sets of data and in extracting physical information. It is worth exploring the applicability of machine learning in research of $\alpha$-particle preformation factor.  

The aim of this work is to explore whether machine learning can be used in the extraction of preformation factor. Here, we incorporate an autoencoder into the traditional multi-layer perceptron neural network to estimate the $\alpha$-particle preformation factors in Sec. \ref{THEORETICAL-FRAMEWORK}. This attempt was conducted under the generalized liquid drop model (GLDM). In Sec. \ref{RESULTS}, adjustment of a free parameter $\beta$ helped us determine the best MLP + AE method. The MLP + AE method captures the relationship between $\alpha$-particle preformation factors and decay energy, and reflects the nuclear structure information of the closed-shell structure and odd-even staggering effect of the nucleus. Furthermore, 41 nuclei in NUBASE2020 \cite{WOS:000625627500001} are used to estimate the extrapolation ability of ML, and the accuracy of the theoretical $\alpha$-decay half-life is greatly improved after taking into account the $\alpha$ formation amplitude. Finally, MLP + AE method is used to calculate the $\alpha$-decay preformation factors of $Z = $119 and 120 isotope chains and the corresponding predicted half-lives are given. The summary will be given in Sec. \ref{SUMMARY}.

\section{THEORETICAL FRAMEWORK}\label{THEORETICAL-FRAMEWORK}
\subsection{The extraction of experimental preformation factors $P_\alpha^\text{exp}$ under the framework of GLDM}

In $\alpha$-decay theory, the decay constant can be written as,
\begin{align}
    \lambda=P_{\alpha}vP,
\end{align}
where $P_\alpha$, $v$ and $P$ denote the $\alpha$-particle preformation factor, the assault frequency, and the barrier penetration probability respectively. The starting point for this work is the relationship between the $\alpha$-decay constant $\lambda$ and the decay half-life $T_{1/2}$, $\lambda=\frac{\ln 2}{T_{1/2}}$. Combining the above two relations and the experimental half-life $T_{1/2}^\text{exp}$, we can get,
\begin{align}
    \lambda_\text{exp}=\frac{\ln 2}{T_{1/2}^\text{exp}}=P_{\alpha}^\text{exp}vP,\\
    \lambda_\text{cal}=\frac{\ln 2}{T_{1/2}^\text{cal}}=P_{\alpha}^\text{cal}vP.
\end{align}
Now we assume that we are in a unified decay theory, and fix the theoretical preformation factor $P_{\alpha}^\text{cal}$ to 1, we can extract the experimental preformation factor as,
\begin{align}
P_{\alpha}^\text{exp}=\frac{T_{1/2}^\text{cal}}{T_{1/2}^\text{exp}}.\label{Pexp}
\end{align}

In this work, we choose the generalized liquid drop model (GLDM) as this unified decay theory, since it can describe the fusion and fission of atomic nuclei well, especially $\alpha$ decay, hence can well support the progress of this work. The assault frequency $\nu=\frac{1}{2R_0}\sqrt{\frac{2E_\alpha}{M_\alpha}}$, where $R_{0}$ is the radius of the parent nucleus given by $R_0=\begin{pmatrix}1.28A^{1/3}-0.76+0.8A^{-1/3}\end{pmatrix}\mathrm{fm}$ and $A$ is the mass number of the parent nucleus, $E_\alpha$ is the energy of the $\alpha$-particle, corrected for recoil, and $M_\alpha$ being its mass. The barrier penetrating probability $P$ is calculated approximately using the Wentzel-Kramers-Brillouin (WKB) method as
\begin{equation}
  P=\exp\left[-\frac{2}{\hbar}\int_{r_{\mathrm{in}}}^{r_{\mathrm{out}}}\sqrt{2B(r)(E(r)-E(\mathrm{sphere}))}dr\right],\label{2}
\end{equation}

where $r_{\mathrm{in}}$ and $r_{\mathrm{out}}$ are the classical turning points which satisfy the conditions 
$r_{\mathrm{in}}=R_{1}+R_{2}$ and 
$E(r_{\mathrm{out}})=Q_{\alpha}$, where $R_{1}$ is the radii of the daughter nuclei and $R_{2}$ is the radii of $\alpha$ nuclei. $B(r) = \mu$ represents the reduced mass. The total interaction potential $E$ in GLDM includes the volume $E_V$, surface $E_S$, Coulomb $E_C$, proximity $E_\mathrm{Prox}$ and the centrifugal $E_l$ energy in turn \cite{WOS:A1985AUN4700008},
\begin{equation}
E=E_V+E_S+E_C+E_\mathrm{Prox}+E_l.\label{3}
\end{equation}
For the detailed expressions of energy in the process of the mother nucleus gradually evolving into daughter nuclei and $\alpha$-particle, please refer to Ref.~\cite{WOS:A1985AUN4700008}. The quasimolecular shape mechanism is used to describe the decay and fusion processes of atomic nuclei in the GLDM, which form a neck as they evolve between one-body and two-body. The emergence of the neck requires considering a proximity energy $E_\mathrm{Prox}$ to optimize the barrier. For a detailed explanation of the proximity energy, see Ref.~\cite{WOS:A1985AUN4700008}. The centrifugal barrier is,
\begin{align}
    E_l=\frac{\hbar^2l(l+1)}{2\mu r^2},
\end{align}
$l$ is the angular momentum carried by the $\alpha$ particle.
\subsection{A hybrid neural network method of multilayer perceptron and autoencoder}
\begin{figure}[htbp]
  \centering 
  \includegraphics[width=0.52\textwidth]{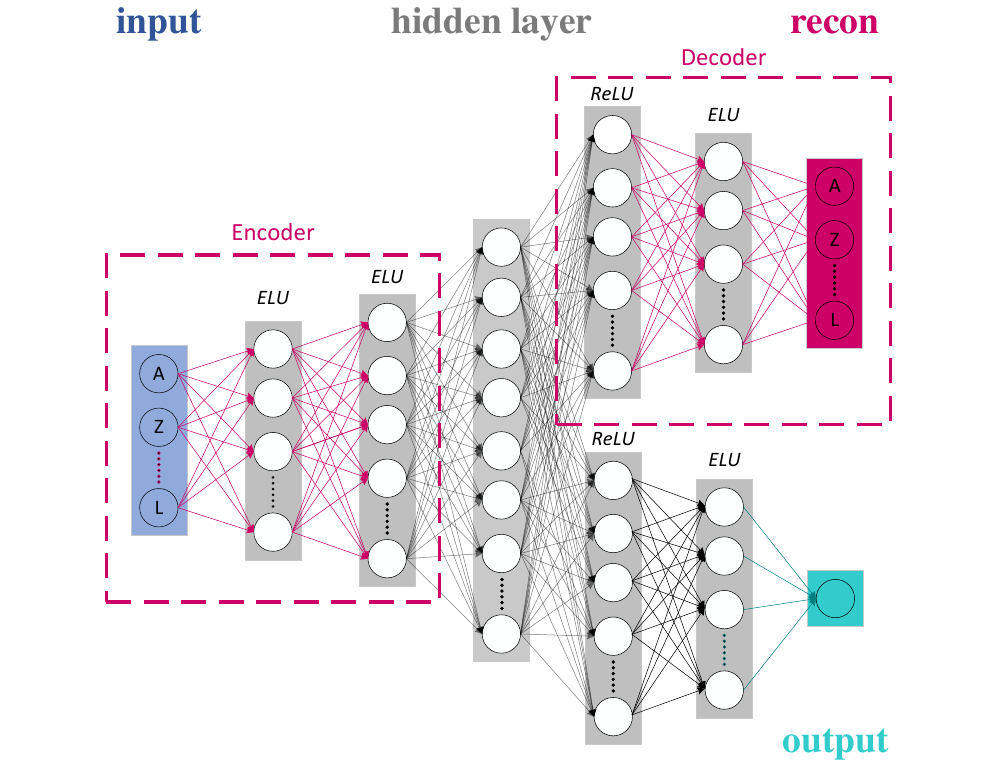}
  \caption{Diagram of the MLP + AE neural network and the autoencoder part is outlined with a dashed line. The number of hidden neural units from left to right is 16, 32, 64, 32, 16. The activation function between layers is shown in the figure.}\label{network}
  \end{figure}
The atomic nucleus is a complex quantum many-body system, and $\alpha$-decay within it is a multi-dimensional decay process involving complex interactions, as well as known and unknown physical factors. Although the $\alpha$-decay half-lives of more than 600 nuclei have been measured experimentally, each nucleus has its own unique characteristics and properties. Therefore, the precise extraction of the half-life and preformation factor from these scarce existing data still faces challenges, especially the prediction and extrapolation to unknown regions based on existing knowledge and physical quantities.

To tackle this challenge, we propose a neural network method that is a hybrid of multi-layer perceptron (MLP) and autoencoder (AE). In general, multilayer perceptron  \cite{WOS:A1994NW17200009} consists of the input layer, the intermediate layer, and the output layer. The data-flow between adjacent layers in the feed-forward neural can be expressed as follows:
\begin{equation}
h_i(x)=\sigma\left(\sum_jw_{ij}x_j+b_i\right),\label{6}
\end{equation}
where \(h_{i}\) is the value of the \(i\)th neuron in the next layer, \(x_{j}\) is the value of the \(j\)th neuron in the previous layer, and \(\sigma\) is the activation function.
\(w_{ij}\) and \(b_i\) are the network weight and the bias, which will be optimized during training. We improve the MLP model based on the architecture of autoencoders and incorporating the reconstruction error \cite{WOS:000237698100002}. 
This approach aims to enhance data representation and improve the generalization performance when encountering small sample sizes. As shown in Fig. \ref{network}, the structure of the MLP + AE neural network is symmetrical, and the number of neurons in the hidden layer from left to right is 16, 32, 64, 32, 16. Unlike the traditional neural network structure, the MLP + AE method has two sets of outputs, one is the reconstruction of the input features $x_\text{recon}$, and the other is the prediction of the label $P^\text{pred}_{\alpha}$.
Therefore, the loss function is modified as follows:
\begin{equation}
  \mathrm{Loss}=\beta\cdot \mathrm{Loss}_\text{recon}+(1-\beta)\cdot \mathrm{Loss}_\text{label},\label{7}
\end{equation}
where $\beta$ is a tunable parameter. $\beta=0$ means that the MLP + AE method returns to the general MLP neural network.
$\text{Loss}_\text{recon}$ and $\text{Loss}_\text{label}$ are given by,
\begin{align}
  \text{Loss}_\text{recon}&=\frac{1}{nm}\sum_{i=1}^{n}\sum_{j=1}^m\left(x_{ij}-x_{ij}^\text{recon}\right)^2, \\
  \text{Loss}_\text{label}&=\frac{1}{n}\sum_{i=1}^n \left(P^\text{exp}_{\alpha i}-P^\text{pred}_{\alpha i}\right)^2, 
  \end{align} 
where \(x_{ij}\) is the input physical feature, $n$ is the total number of data and $m$ is the number of physical quantity.

The MLP + AE neural network is trained using the Adam optimization algorithm \cite{kingma2014adam} and uses batch normalization to speed up training and avoid the disappearance and explosion of gradients \cite{WOS:000684115800048}. Due to the small size of the dataset, the model is trained in 100 epochs only and regularization of L2 is added to reduce overfitting \cite{WOS:000724320000009}. In order to achieve better training results and reduce the impact of smaller datasets, it is necessary to preprocess the raw data. Therefore, the input layer consists of eight neurons, corresponding to eight physical quantity inputs, including mass \(A\), proton \(Z\), neutron \(N\), decay energy \(Q_{\alpha}\), and the minimum angular momentum $l$, as well as \(x_{1}\), \(x_{2}\), \(x_{3}\), which are given by the following equation,
\begin{equation}
x_{1}=A_{1}^{1/6}+A_{2}^{1/6},~~~~~~
  x_{2}=\frac{N}{\sqrt{Q_{\alpha}}},~~~~~~~
  x_{3}=\sqrt{l(l+1)}.\label{12}
\end{equation}
where $A_{1}$ and $A_{2}$ are the mass numbers of $\alpha$-particle and daughter nuclei. Because data partitioning also has a significant impact on the performance of the model, we do 1000 times tenfold cross validation.
The average of the 1000 results is chosen as the average performance of the model. Two activation functions, ReLU and ELU \cite{WOS:000843491800008} are used in the network structure, and the details are shown in Fig. \ref{network}.

\section{RESULTS}\label{RESULTS}
\subsection{Effect of $\beta$ on $\alpha$-particle preformation factor with the hybrid neural network method of MLP + AE}
Machine learning in the field of nuclear physics usually includes direct learning and indirect learning. Direct learning is to learn the experimental values of the physical targets, so that one can test the learning ability of machine learning for the physical targets. While indirect learning is learning the difference between the experimental value and the theoretical value of the physical target, and then using it for the optimization of the theoretical calculation. In this work, we calculate the \(\alpha\)-decay half-lives using the GLDM with \(P_{\alpha}=1\), and extract the experimental preformation factor of 535 nuclei according to Eq. \eqref{Pexp}, including 159 even-even nuclei, 295 odd-$A$ nuclei, and 81 odd-odd nuclei. The experimental $\alpha$-decay half-lives and spins in the dataset are derived from the evaluated nuclear properties table the NUBASE2016 \cite{WOS:000407994800001}. 
The $\alpha$-decay energies are derived from evaluated atomic mass table AME2016 \cite{WOS:000407994800002,WOS:000407994800003}. Here, the MLP model is improved by introducing the parameter \(\beta\) based on the autoencoder, and we use indirect learning, that is, the extracted preformed factors are obtained indirectly from the GLDM and then MLP + AE learns them. In order to assess the consistency between of the $\alpha$-particle preformation factor estimated by the MLP + AE method and the extracted $\alpha$-particle preformation factor, the root-mean-square deviation ($\sigma_\text{RMS}$) is used to quantify the performance of the model, which is defined as,
\begin{equation}
\sigma_\text{RMS}=\sqrt{\frac1n\sum_{i=1}^{n}(\log_{10}P_{\alpha i}^{\mathrm{MLP+AE}}-\log_{10}P_{\alpha i}^{\mathrm{exp}})^2}.\label{rms}
\end{equation}
Here, $\log_{10}P_{\alpha i}^{\mathrm{MLP+AE}}$ is the logarithm of preformation factor given after learning by the MLP + AE method, $\log_{10}P_{\alpha i}^{\mathrm{exp}}$ is the logarithm of the experimental preformation factor extracted from the GLDM model and $N = 535$.

\begin{figure}[htbp]
	\centering
\includegraphics[width=0.47\textwidth]{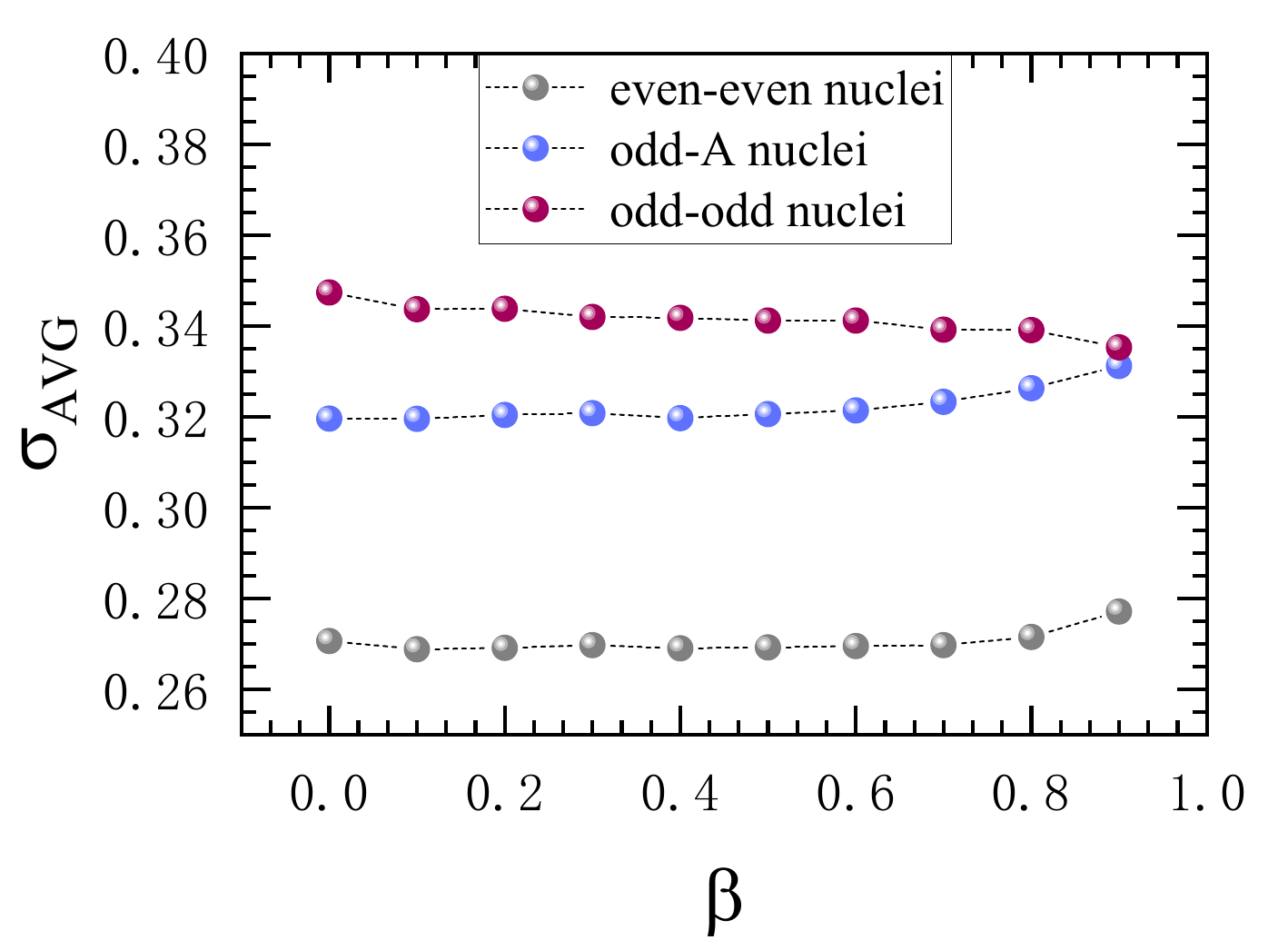}
	\caption{The average $\sigma_{\mathrm{AVG}}$ of the even-even, odd-$A$, and odd-odd nuclei as a function of $\beta$ value.}
  \label{beta}
\end{figure}

\begin{table*}[htbp]
  \renewcommand{\arraystretch}{1}
  \setlength\tabcolsep{4pt}
    \centering
    \caption{The average effect \(\sigma_\text{AVG}\) of  the model with different values of $\beta$ for different types of nuclei.}
    \begin{tabular}{lcccccccccc}
      \hline\hline
    ~~~~~~Type &\multicolumn{9}{c}{\(\sigma_\text{AVG}\)}\\ \cline{1-11}
    ~~~~~~~~~$\beta$&0&0.1&0.2&0.3&0.4&0.5&0.6
    &0.7&0.8&0.9\\
    ~~even-even nuclei~~&~~0.2706~~ &~~0.2688~~&~~0.2691~~&~~0.2697~~&~~0.2690~~&~~0.2692~~&~~0.2695~~&~~0.2697~~&~~0.2715~~&~~0.2771~~\\
    ~~odd-$A$ nuclei~~    &~0.3196~ &~0.31955~&~~0.3204~~&~~0.3208~~&~~0.3197~~&~~0.3206~~&~~0.3214~~&~~0.3233~~&~~0.3263~~&~~0.3312~~\\
    ~~odd-odd nuclei~~  &~~0.3474~~ &~~0.3437~~&~~0.3438~~&~~0.3420~~&~~0.3418~~&~~0.3412~~&~~0.3412~~&~~0.3392~~&~~0.3391~~&~~0.3353~~\\
    \hline\hline
    \label{TABLE1}
    \end{tabular}
  \end{table*}
  
A $\mathcal{K}$-fold cross-validation method is introduced to avoid overfitting and improve the generalization performance of MLP + AE. In $C$-times of cross-validation, the average effect of the model is defined as,
\begin{equation}
  \sigma_{\mathrm{AVG}}=\frac{1}{C\mathcal{K}}\sum_{j=1}^{C}\left(\sum_{i=1}^\mathcal{K}\left(\sigma_{\mathrm{RMS}}\right)_{i}\right)_{j},\label{14}
\end{equation}
where $\mathcal{K}$ is the number of folds, and $C$ is the number of cross-validation runs. The average performance of the model, evaluated through 1000 times of tenfold cross-validation for even-even, odd-$A$, and odd-odd nuclei at various values of $\beta$, is summarized in Table \ref{TABLE1}. The variation of $\sigma_{\mathrm{AVG}}$ with $\beta$ is also shown in Fig. \ref{beta}. When $\beta=0.1$, the model exhibits an optimal average performance for even-even and odd-$A$ nuclei. In contrast, at $\beta=0.9$, the model demonstrates a superior average performance for odd-odd nuclei. 
Specifically, for even-even nuclei, the value of $\sigma_{\mathrm{AVG}}$ decreased from 0.2706 to 0.2688.
For odd-$A$ nuclei, the value of $\sigma_{\mathrm{AVG}}$ decreased from 0.31958 to 0.31955. 
Notably, for odd-odd nuclei, the value of $\sigma_{\mathrm{AVG}}$ decreased from 0.3474 to 0.3353 with a decrease of 3.5\%. These results indicate that the improved MLP model has enhanced its ability to estimate preformation factor, particularly for odd-odd nuclei. The introduction of the autoencoder can test the stability of MLP method learning. This cross-validation approach ensures comprehensive coverage of the entire dataset and makes the overall performance of MLP + AE tend to be ideal. In addition, the addition of AE makes the calculation of preformation factors more flexible and reliable.
\subsection{Preformation factor and nuclear structure information}
Figure \ref{beta} in the previous subsection shows how the average effect $\sigma_{\mathrm{AVG}}$ of different nuclei varies with $\beta$. For even-even nuclei and odd-$A$ nuclei, the optimal $\beta$ value is 0.1, and for odd-odd nuclei, the optimal $\beta$ value is 0.9. After determining the optimal value of $\beta$ based on the optimal value of average performance, the one with the smallest value of $\sigma_{\mathrm{RMS}}$ out of 1000 times tenfold cross-validations is chosen as the final result. Then the $\alpha$-particle preformation factors can be estimated for all 535 nuclei. 

\begin{figure}[htbp]
  \centering
  \includegraphics[width=0.5\textwidth]{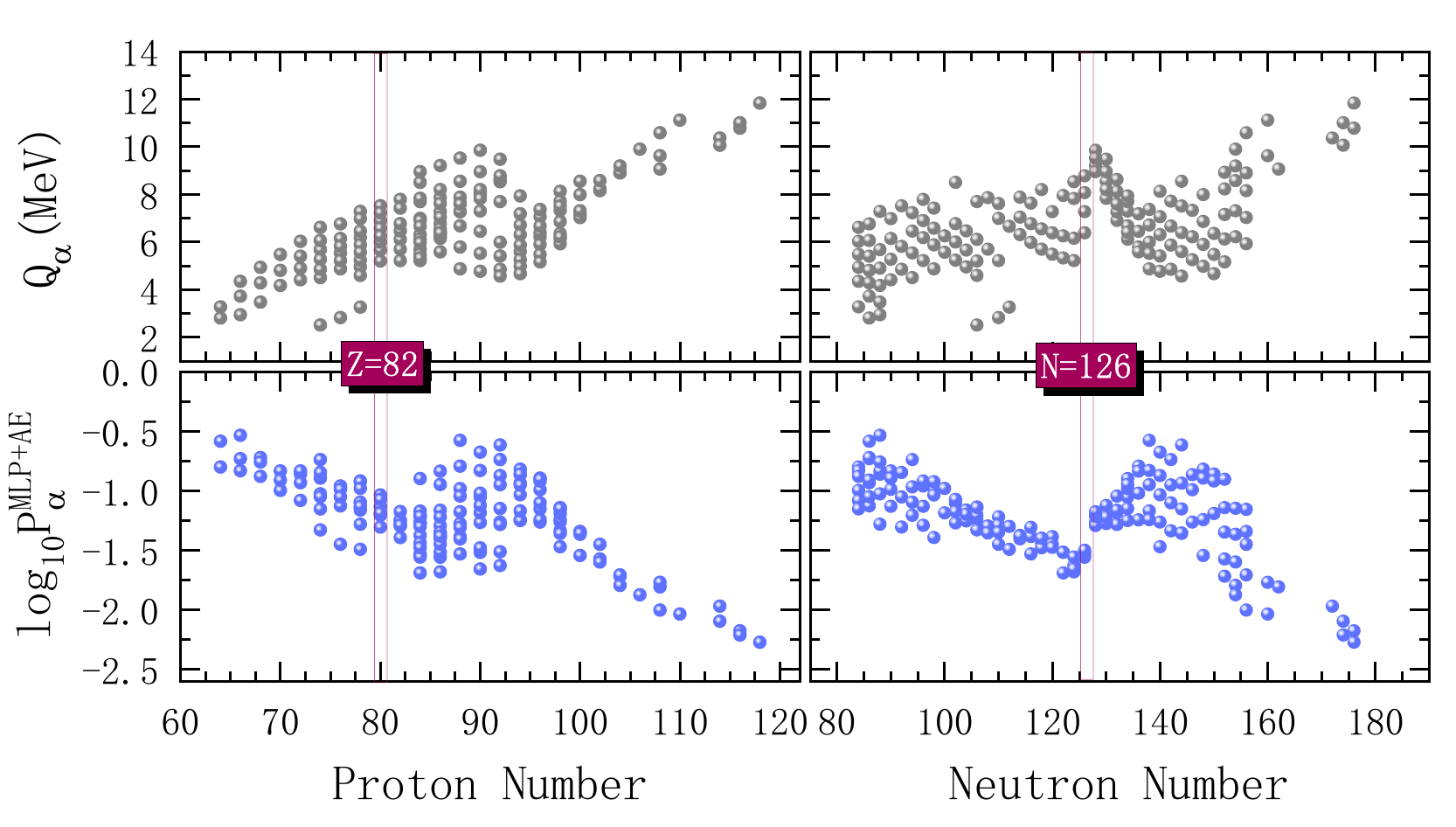}
	\caption{Upper: The experimental decay energy $Q_\alpha$ (in MeV) for even-even nuclei as a function of the proton number (left) and the neutron number (right). Lower: The logarithmic values of preformation factors with MLP + AE as a function of $Z$ (left) and $N$ (right).}
  \label{82126}
\end{figure}

The logarithmic values of preformation factor with MLP + AE method and the corresponding $\alpha$-decay energy $Q_\alpha$ (in MeV), a key input to the neural network, as functions of proton number $Z$ and neutron number $N$ are shown in Fig. \ref{82126}. Before $Z$=82 and $N$=126, as the number of protons and neutrons increases, the decay energy $Q_\alpha$ increases gently, while the logarithm of the preformation factor decreases. After $Z$=82 and $N$=126, both $Q_\alpha$ and $\text{log}_{10}P_\alpha^\text{MLP+AE}$ change significantly. As for the preformation factor, after $Z$ crossing $Z = 82$ shell closure, $\text{log}_{10}P_\alpha^\text{MLP+AE}$ increase at first and then drop down with $Z$ approaching the next proton full shell. Similar scenes appear in the region of $N>$ 126. Obviously, the changing trends of $\text{log}_{10}P_\alpha^\text{MLP+AE}$ on both sides of the $Z$=82 and $N$=126 closed shells are different. This indicates that the preformation factor reflects the closed-shell effect of the nucleus. In this way, preformation factors can also become a positive signal for exploring the stability of superheavy nuclei. The same as the results in Ref.\cite{WOS:000647421500074}, we find that the trend of $\text{log}_{10}P_\alpha^\text{MLP+AE}$ can also be concluded that the shell effect of ($Z$=82, $N$=126) is stronger than that of ($Z$=50, $N$=82). In addition, the logarithmic preformation factor and decay energy show very regular and symmetrical trends, which is also an important basis for the empirical formula proposed in Ref.~\cite{WOS:000647421500074}.

\begin{figure*}[hbtp]
	\centering
  \includegraphics[width=0.86\textwidth]{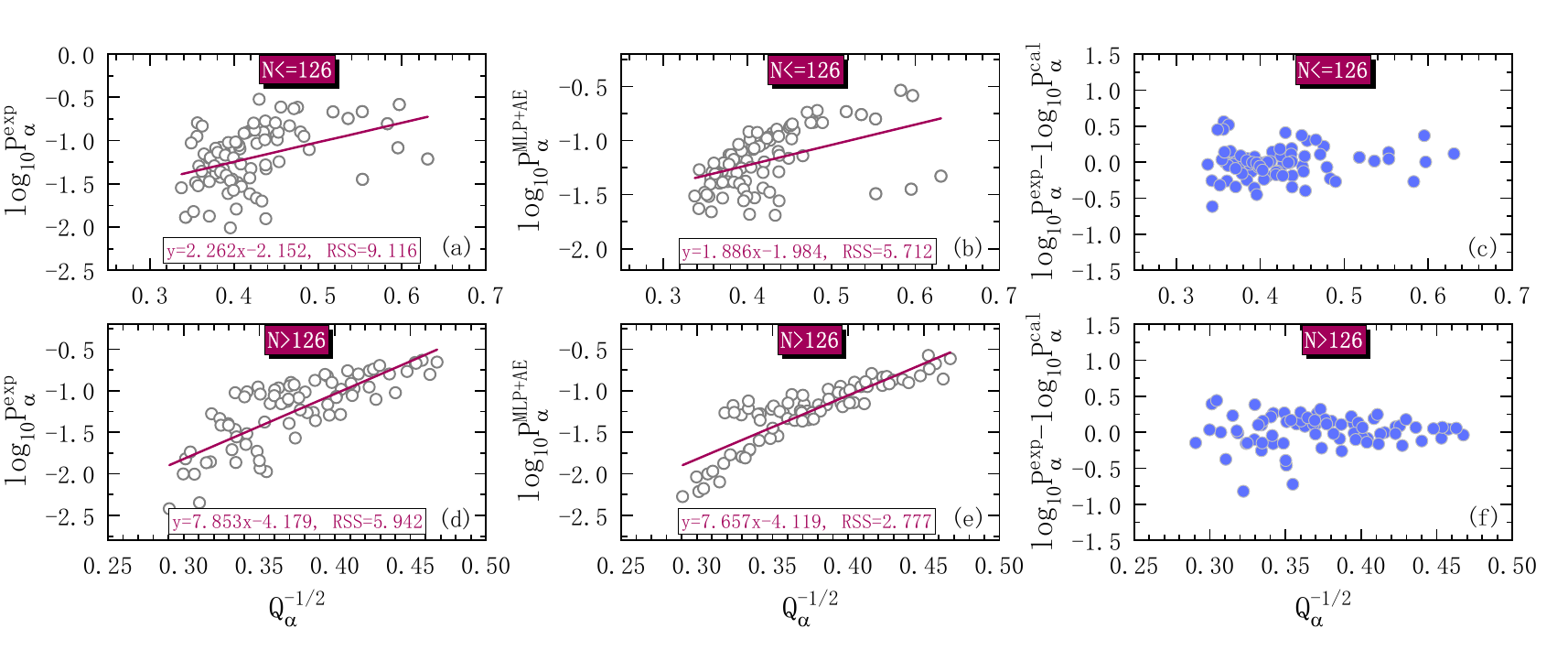}
  \caption{From left to right: the extracted logarithm of the experimental preformation factor, the value calculated by MLP + AE method, and logarithmic differences between the extracted and calculated of $\alpha$-particle preformation factors as a function of $Q_\alpha^{-1/2}$ (in MeV) for even-even nuclei. The cases for $N \leqslant 126$ in upper column and $N > 126$ in the bottom column. The results of linear fitting and the RSS value for evaluating the fitting effect are given.}
  \label{P-Q}
\end{figure*}

\begin{figure}[htbp]
	\centering
  \includegraphics[width=0.4594\textwidth]{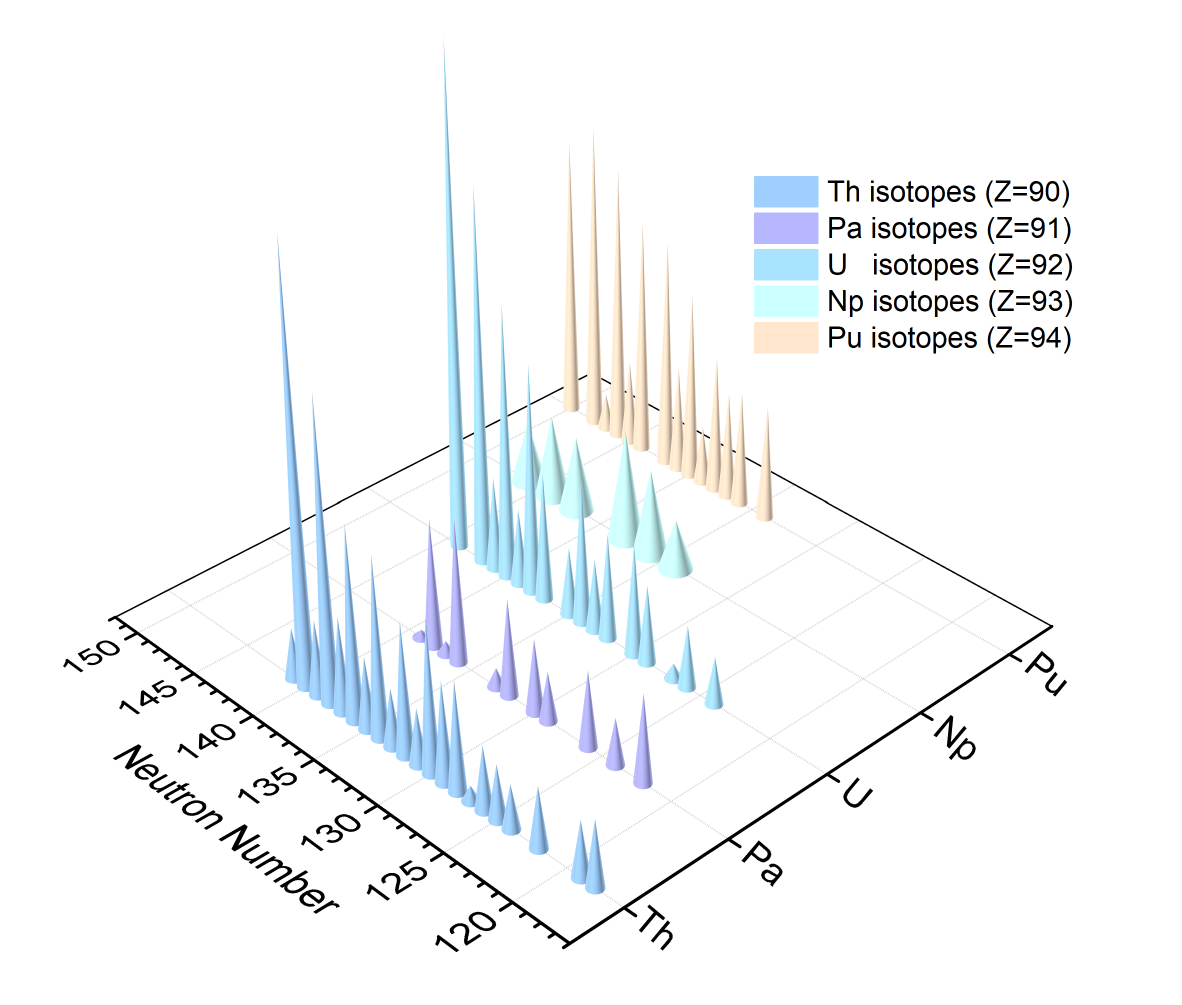}
    \includegraphics[width=0.4594\textwidth]{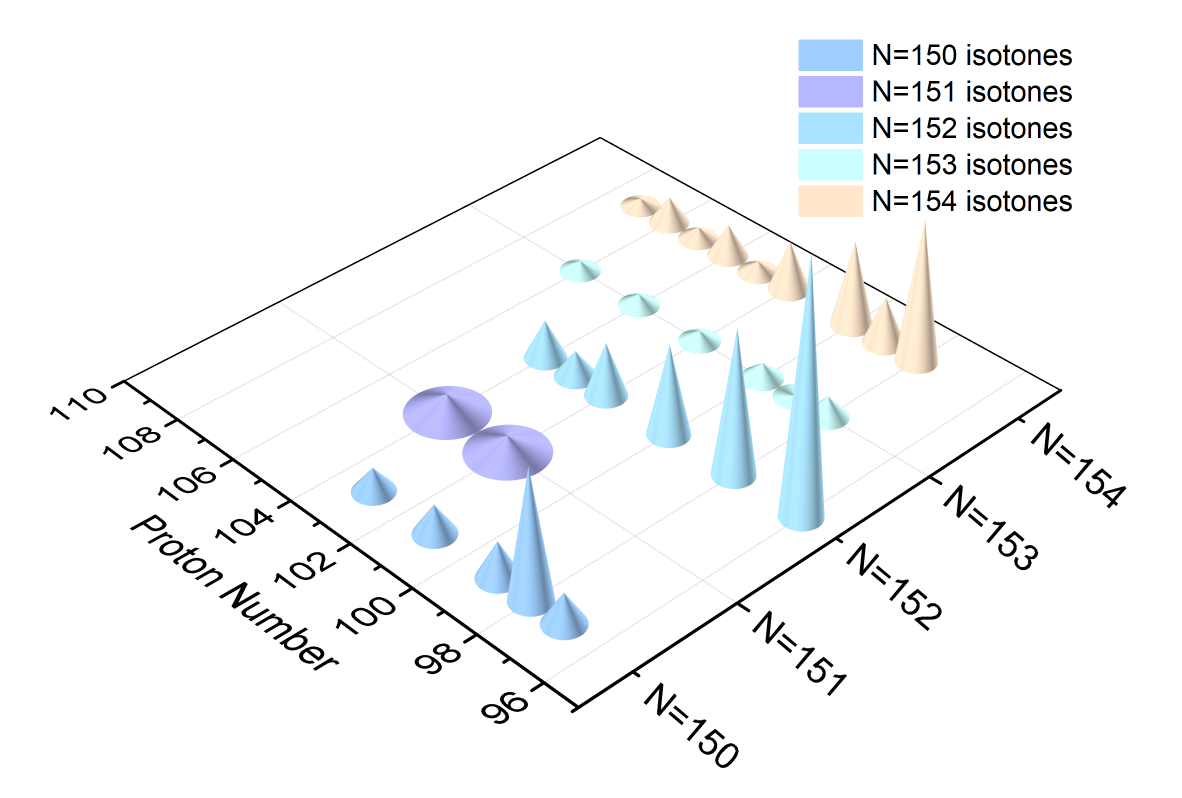}
  \caption{Upper: The calculated $\alpha$-particle preformation factors with the MLP + AE method as a function of neutron number for the Th, Pa, U, Np and Pu isotope chains; Lower: the $P_\alpha^\text{MLP+AE}$ for four isotones, $N = 150, 151, 152, 153$ and 154, but as a function of proton number.}
  \label{four-isotopes}
\end{figure}

There is an approximate linear relationship between the logarithm of the experimental preformation factor $\text{log}_{10}P_\alpha^\text{exp}$ and the square root of the decay energy ratio $Q_\alpha^{-1/2}$, in the areas of $N\leqslant$ 126 and $N>$ 126 respectively \cite{WOS:000647421500074}. It should be necessary to verify here whether the MLP + AE method can capture this physical phenomenon in this work. Figures \ref{P-Q}(a) and (b) show the relationship between the experimental value $\text{log}_{10}P_\alpha^\text{exp}$ and $Q_\alpha^{-1/2}$, and $\text{log}_{10}P_\alpha^\text{exp}$ and $Q_\alpha^{-1/2}$ in the area of $N\leqslant$ 126, respectively; (d) and (e) correspond to the region $N>$ 126, and the corresponding linear fit and residual sum of squares (RSS) are shown in the figure to quantitatively evaluate the fitting effect; (c) and (f) show the difference between the logarithm of the experimental preformation factor and the MLP + AE value in the two regions as $Q_\alpha^{-1/2}$ changes. We can see that there is an obvious linear relationship between the two physical quantities, and the fitting parameters also support this conclusion. The difference values in Fig. \ref{P-Q} are basically distributed around zero, and there is no obvious linear relationship between the value of ($\text{log}_{10}P_\alpha^\text{exp}-\text{log}_{10}P_\alpha^\text{MLP+AE}$) with $Q_\alpha^{-1/2}$. This shows that the MLP + AE neural network successfully learned the inherent rules between these two quantities, and the learning is effective. This result further supports that the Geiger-Nuttall law can not only describe $\alpha$-decay half-lives but also deal with $\alpha$-particle preformation factors.
The odd-even staggering effect is a common physical phenomenon in many physical quantities in nuclear physics, such as atomic nucleus mass, one-nucleon and two-nucleon separation energies, decay energy \cite{Yang1}, half-life \cite{Yang2}, etc. Whether this kind of phenomenon also exists in the $\alpha$-particle preformation factors deserves to be studied. Figure \ref{four-isotopes} plots the calculated preformation factors for the Th, Pa, U, Np and Pu isotope chains, using the MLP + AE method. The results are satisfactory and the odd-even staggering effect also appear in the $P_\alpha^\text{MLP+AE}$. For the three even-$Z$ isotope chains of Th, U and Pu, the preformation factor corresponding to odd-$N$ nuclei is smaller than that of the adjacent even-$N$ nuclei. This shows that unpaired neutron inhibits the formation of $\alpha$-particle on the surface of the parent nuclei. Therefore, the value of $P_\alpha^\text{MLP+AE}$ is relatively small. In the same way, compared with the Pa isotope chain of an odd-$Z$ and the Th isotope chain of an even-$Z$, when the neutron number $N$ is same, the $P_\alpha^\text{MLP+AE}$ value of the Pa isotope chain is basically smaller than the corresponding value of the Th isotope chain, which means that unpaired proton also prevents the formation of $\alpha$ particles. In general, the Th isotope chain with $Z = 90$ is larger than the adjacent Pa isotope chain ($Z = 91$). The preformation factor of the Pu isotope chain ($Z = 94$) is greater than that of the adjacent Np isotope chain with $Z = 93$. When the neutron number is 144, the preformation factor of the U isotope is larger than that of the Th and Pu isotope chains. Combined with the conclusion in Fig. \ref{82126}, the preformation factor first increases with increasing the number of protons after passing the magic number of $Z = 82$. It increases, reaching the maximum value at about $Z = 92$, and then gradually decreases after 92. Figure \ref{four-isotopes} also shows the preformation factors for the $N = 150, 151, 153$ and 154 isotone chains. It can be seen that these isotones chains also show an obvious odd-even staggering effect. In short, unpaired nuclei reduce the probability of $\alpha$-particles forming on the surface of the parent nucleus. In other words, trends and changes in preformation factors can give information about the structure of the nucleus.
In previous studies, the $\alpha$-particle preformation factors have typically been estimated using empirical formulas. Empirical formulations from Refs.~\cite{WOS:000647421500074,WOS:000606025400001} are selected for comparison with the calculation of the MLP + AE neural network.
The values of $\sigma_{\mathrm{RMS}}$ of them are presented in Table \ref{TABLE2}. Comparing the root mean square error of our calculation with two sets of empirical formulas, we find that the results of machine learning are comparable to or even better than the results of the empirical formulas. Specifically, when compared to Ref.~\cite{WOS:000606025400001}, the $\sigma_{\mathrm{RMS}}$ values for the four types of nuclei decrease from 0.280, 0.417, 0.359, 0.397 to 0.218, 0.296, 0.282, 0.273. Compared with Ref.~\cite{WOS:000647421500074}, the corresponding value of $\sigma_{\mathrm{RMS}}$ for the four types of nuclei decreases from 0.219, 0.321, 0.306, 0.340 to 0.218, 0.296, 0.282, 0.273.
The overall $\sigma_{\mathrm{RMS}}$ decreased from 0.365 and 0.293 to 0.268, with a decrease of 26.6\% and 8.5\%, respectively.

\begin{table}[htbp]
  \renewcommand{\arraystretch}{1.1}
  \setlength\tabcolsep{10pt}
    \caption{Comparison of the values of $\sigma_{\mathrm{RMS}}$ for the improved MLP model and the empirical formulation.}
    \centering
    \begin{tabular}{cccc}
      \hline\hline
   nuclei~type &MLP + AE&Ref.~\cite{WOS:000606025400001} &Ref.~\cite{WOS:000647421500074}\\ \hline
    ~~even-even nuclei  &~~0.218~~ &~~0.280~~&~~0.219~~\\
    ~~even-odd nuclei  &~~0.296~~ &~~0.417~~&~~0.321~~\\
    ~~odd-even nuclei  &~~0.282~~ &~~0.359~~&~~0.306~~\\
    ~~odd-odd nuclei  &~~0.273~~ &~~0.397~~&~~0.340~~\\
    ~~all  &~~0.268~~ &~~0.365~~&~~0.293~~\\
    \hline\hline
    \label{TABLE2}
    \end{tabular}
  \end{table}


\subsection{Extrapolation capability of MLP + AE neural network}\label{Extrapolation}
The half-life is a physical quantity that can be measured directly in experiments. The half-life that takes into account the probability of $\alpha$-particle formation should be more consistent with the experimental value. In order to test the performance of the preformation factor calculated by MLP + AE method. We first calculated the preformation factors of the 41 newly added nuclei ($Z>$ 104) in NUBASE2020 relative to the 2016 mass table and compared them with the results calculated by the empirical formula \cite{WOS:000647421500074}. The decay energies are derived from the evaluated atomic mass table AME2020. The detailed results are listed in Table \ref{TBALE3}. The preformation factors calculated by the MLP + AE neural network are basically consistent with the results calculated by the empirical formula, indicating that the MLP + AE method can be used for the extrapolation and prediction of preformation factors of nuclei in unknown nuclear regions. 
\begin{figure}[htbp]
  \centering
\includegraphics[width=0.5\textwidth]{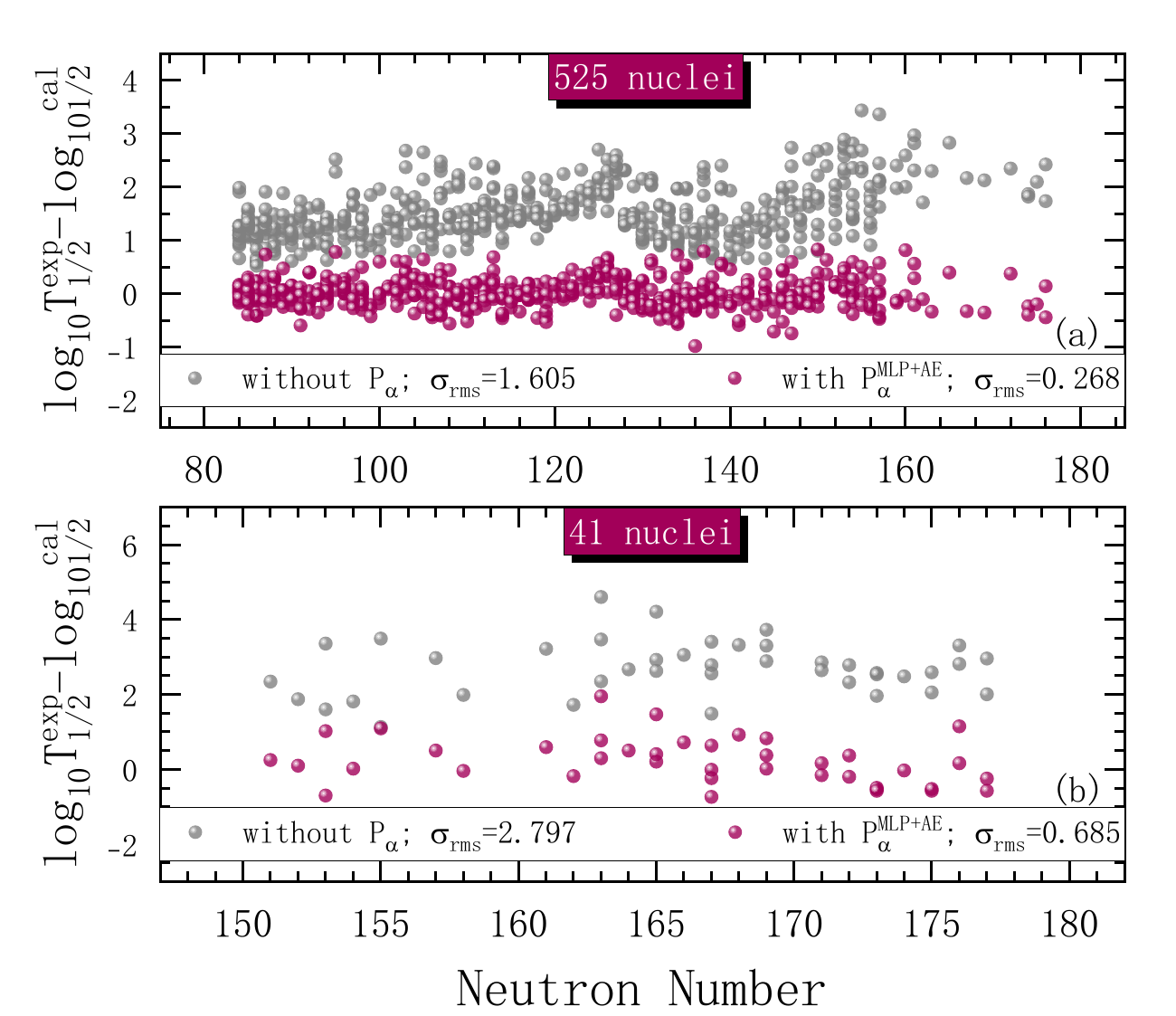}
  \caption{Upper: The difference between the experimental half-life and the theoretical data under the GLDM framework for 535 nuclei, including those with and without considering $P_\alpha^\text{MLP+AE}$. Lower: For 41 newly added nuclei.}
  \label{D-life}
\end{figure}

Furthermore, we calculate the half-life after considering the preformation factor and compare it with the experimental value. Figure \ref{D-life} shows the difference between the experimental half-life and the theoretical data under the GLDM framework for 535 nuclei and 41 nuclei, including those with and without considering $P_\alpha^\text{MLP+AE}$. It is obvious that after considering the preformation factor, the half-life rms for 535 nuclei dropped from 1.605 to 0.268, a decrease of 83$\%$, and for 41 test nuclei, the rms dropped from 2.797 to 0.685, a decrease of 75$\%$. The accuracy of the half-life further proves the reliability of the preformation factor calculated by the MLP + AE approach.

\begin{table*}[htp]
   \centering
     \caption{Comparison of the predictions of the improved MLP model with the empirical formulation from Ref.\cite{WOS:000647421500074} 
     of the $\alpha$-particle preformation factors for 41 new nuclei taken from NUBASE2020\cite{WOS:000625627500001}.
     $P_{\alpha}^\text{Deng}$ is $\alpha$-particle preformation factor calculated by the empirical formula from Ref.\cite{WOS:000647421500074}. $P_{\alpha}^\text{MLP+AE}$ is $\alpha$-particle preformation factor calculated by the improved MLP model.
     }
     \begin{tabular}{ccccccccccc}
       \hline\hline
     $\alpha$ transition&\(Q_{\alpha}\)&$j_{p}^{\pi}\rightarrow j_{d}^{\pi}$   &$l_\text{min}$&$\log_{10}P_{\alpha}^\text{Deng}$&$P_{\alpha}^\text{Deng}$&$\log_{10}P_{\alpha}^\text{MLP+AE}$&$P_{\alpha}^\text{MLP+AE}$&$\log_{10}T_{1/2}^\text{exp}$&$\log_{10}T_{1/2}^\text{GLDM}$&$\log_{10}T_{1/2}^\text{GLDM+$P_\alpha$}$\\ \hline
     ${}^{255}\mathrm{Rf}\to{}^{251}\mathrm{No}$&9.06  &$7/2^+\to9/2^-$&1&-2.00&$9.91\times 10^{-3}$&-1.88&$1.33\times 10^{-2}$&	0.49 	&	-1.86 	&	0.24 \\
     ${}^{256}\mathrm{Rf}\to{}^{252}\mathrm{No}$&8.93  &$0  ^+\to0  ^+$&0&-1.67&$2.13\times 10^{-2}$&-1.78&$1.66\times 10^{-2}$&	0.33 	&	-1.55 	&	0.23 	\\
     ${}^{257}\mathrm{Rf}\to{}^{253}\mathrm{No}$&9.08&$9/2^-\to1/2^+$&5&-2.50&$3.17\times 10^{-3}$&-2.30&$4.99\times 10^{-3}$&	0.75 	&	-0.85 	&	1.45 	\\
     ${}^{258}\mathrm{Rf}\to{}^{254}\mathrm{No}$&9.20 &$0  ^+\to0  ^+$&0&-1.71&$1.94\times 10^{-2}$&-1.80&$1.59\times 10^{-2}$&	-0.59 	&	-2.41 	&	-0.61 	\\
     ${}^{269}\mathrm{Sg}\to{}^{265}\mathrm{Rf}$&8.58 &$0  ^+\to0  ^+$&0&-2.16&$6.96\times 10^{-3}$&-2.06&$8.73\times 10^{-3}$&	2.48 	&	0.12 	&	2.18 	\\
     ${}^{260}\mathrm{Bh}\to{}^{256}\mathrm{Db}$&10.40 &$0  ^+\to0  ^+$&0&-2.73&$1.85\times 10^{-3}$&-2.34&$4.54\times 10^{-3}$&	-1.39 	&	-4.75 	&	-2.41 	\\
     ${}^{262}\mathrm{Bh}\to{}^{258}\mathrm{Db}$&10.32&$0  ^+\to0  ^+$&0&-2.74&$1.84\times 10^{-3}$&-2.40&$4.01\times 10^{-3}$&	-1.08 	&	-4.57 	&	-2.17 	\\
     ${}^{270}\mathrm{Bh}\to{}^{266}\mathrm{Db}$&9.06&$0  ^+\to0  ^+$&0&-2.64&$2.29\times 10^{-3}$&-2.69&$2.02\times 10^{-3}$&	2.36 	&	-1.11 	&	1.58 	\\
     ${}^{271}\mathrm{Bh}\to{}^{267}\mathrm{Db}$&9.42&$0  ^+\to0  ^+$&0&-2.74&$1.80\times 10^{-3}$&-2.17&$6.74\times 10^{-3}$&	0.46 	&	-2.21 	&	-0.04 	\\
     ${}^{272}\mathrm{Bh}\to{}^{268}\mathrm{Db}$&9.30 &$0  ^+\to0  ^+$&0&-2.81&$1.56\times 10^{-3}$&-2.72&$1.91\times 10^{-3}$&	1.05 	&	-1.87 	&	0.85 	\\
     ${}^{274}\mathrm{Bh}\to{}^{270}\mathrm{Db}$&8.94&$0  ^+\to0  ^+$&0&-2.92&$1.22\times 10^{-3}$&-2.80&$1.59\times 10^{-3}$&	1.76 	&	-0.81 	&	1.99 	\\
     ${}^{263}\mathrm{Hs}\to{}^{259}\mathrm{Sg}$&10.73&$11/2^-\to3/2^+\#$&5&-2.03&$9.25\times 10^{-3}$&-2.47&$2.69\times 10^{-3}$&	-3.05 	&	-4.18 	&	-1.61 	\\
     ${}^{266}\mathrm{Hs}\to{}^{262}\mathrm{Sg}$&10.35&$0  ^+\to0  ^+$&0&-1.92&$1.20\times 10^{-2}$&-2.03&$9.33\times 10^{-3}$&	-2.40 	&	-4.39 	&	-2.36 	\\
     ${}^{270}\mathrm{Hs}\to{}^{266}\mathrm{Sg}$&9.07&$0  ^+\to0  ^+$&0&-1.80&$1.57\times 10^{-2}$&-1.91&$1.24\times 10^{-2}$&	0.95 	&	-0.77 	&	1.14 	\\
     ${}^{273}\mathrm{Hs}\to{}^{269}\mathrm{Sg}$&9.65&$0  ^+\to0  ^+$&0&-2.20&$6.24\times 10^{-3}$&-2.22&$5.97\times 10^{-3}$&	0.03 	&	-2.60 	&	-0.38 	\\
     ${}^{275}\mathrm{Hs}\to{}^{271}\mathrm{Sg}$&9.45 &$0  ^+\to0  ^+$&0&-2.31&$4.87\times 10^{-3}$&-2.23&$5.94\times 10^{-3}$&	-0.55 	&	-2.04 	&	0.19 	\\
     ${}^{266}\mathrm{Mt}\to{}^{262}\mathrm{Bh}$&11.00&$0  ^+\to0  ^+$&0&-2.64&$2.29\times 10^{-3}$&-2.47&$3.37\times 10^{-3}$&	-2.70 	&	-5.67 	&	-3.20 	\\
     ${}^{275}\mathrm{Mt}\to{}^{271}\mathrm{Bh}$&10.48&$0  ^+\to0  ^+$&0&-2.19&$6.45\times 10^{-3}$&-2.34&$4.59\times 10^{-3}$&	-1.51 	&	-4.57 	&	-2.23 	\\
     ${}^{276}\mathrm{Mt}\to{}^{272}\mathrm{Bh}$&10.10&$0  ^+\to0  ^+$&0&-2.95&$1.12\times 10^{-3}$&-2.78&$1.67\times 10^{-3}$&	-0.15 	&	-3.56 	&	-0.79 	\\
     ${}^{278}\mathrm{Mt}\to{}^{274}\mathrm{Bh}$&9.58&$0  ^+\to0  ^+$&0&-2.85&$1.40\times 10^{-3}$&-2.87&$1.35\times 10^{-3}$&	0.78 	&	-2.11 	&	0.76 	\\
     ${}^{272}\mathrm{Rg}\to{}^{268}\mathrm{Mt}$&11.20&$0  ^+\to0  ^+$&0&-2.68&$2.11\times 10^{-3}$&-2.63&$2.33\times 10^{-3}$&	-2.38 	&	-5.60 	&	-2.97 	\\
     ${}^{274}\mathrm{Rg}\to{}^{270}\mathrm{Mt}$&11.48&$0  ^+\to0  ^+$&0&-2.84&$1.44\times 10^{-3}$&-3.66&$2.20\times 10^{-3}$&	-1.70 	&	-6.31 	&	-3.65 	\\
     ${}^{278}\mathrm{Rg}\to{}^{274}\mathrm{Mt}$&10.85&$0  ^+\to0  ^+$&0&-2.91&$1.22\times 10^{-3}$&-2.80&$1.59\times 10^{-3}$&	-2.10 	&	-4.88 	&	-2.08 	\\
     ${}^{279}\mathrm{Rg}\to{}^{275}\mathrm{Mt}$&10.53&$0  ^+\to0  ^+$&0&-2.36&$4.36\times 10^{-3}$&-2.40&$3.99\times 10^{-3}$&	-0.77 	&	-4.09 	&	-1.69 	\\
     ${}^{280}\mathrm{Rg}\to{}^{276}\mathrm{Mt}$&10.15&$0  ^+\to0  ^+$&0&-3.04&$9.20\times 10^{-4}$&-2.90&$1.25\times 10^{-3}$&	0.63 	&	-3.10 	&	-0.20 	\\
     ${}^{278}\mathrm{Nh}\to{}^{274}\mathrm{Rg}$&11.99&$0  ^+\to0  ^+$&0&-2.76&$1.75\times 10^{-3}$&-2.75&$1.79\times 10^{-3}$&	-2.64 	&	-6.85 	&	-4.10 	\\
     ${}^{282}\mathrm{Nh}\to{}^{268}\mathrm{Rg}$&10.78&$0  ^+\to0  ^+$&0&-2.95&$1.12\times 10^{-3}$&-2.93&$1.17\times 10^{-3}$&	-0.85 	&	-4.16 	&	-1.23 	\\
     ${}^{284}\mathrm{Nh}\to{}^{280}\mathrm{Rg}$&10.28&$0  ^+\to0  ^+$&0&-2.91&$1.23\times 10^{-3}$&-3.02&$9.60\times 10^{-4}$&	-0.01 	&	-2.87 	&	0.14 	\\
     ${}^{285}\mathrm{Nh}\to{}^{281}\mathrm{Rg}$&10.01&$0  ^+\to0  ^+$&0&-2.43&$3.75\times 10^{-3}$&-2.42&$3.80\times 10^{-3}$&	0.66 	&	-2.12 	&	0.30 	\\
     ${}^{286}\mathrm{Nh}\to{}^{282}\mathrm{Rg}$&9.79&$0  ^+\to0  ^+$&0&-2.86&$1.37\times 10^{-3}$&-3.10&$7.88\times 10^{-4}$&	1.08 	&	-1.49 	&	1.61 	\\
     ${}^{285}\mathrm{Fl}\to{}^{281}\mathrm{Cn}$&10.56&$0  ^+\to0  ^+$&0&-2.36&$4.40\times 10^{-3}$&-2.48&$3.29\times 10^{-3}$&	-0.68 	&	-3.32 	&	-0.84 	\\
     ${}^{287}\mathrm{Fl}\to{}^{283}\mathrm{Cn}$&10.17&$0  ^+\to0  ^+$&0&-2.47&$3.41\times 10^{-3}$&-2.46&$3.45\times 10^{-3}$&	-0.29 	&	-2.26 	&	0.20 	\\
     ${}^{290}\mathrm{Fl}\to{}^{286}\mathrm{Cn}$&9.86 &$0  ^+\to0  ^+$&0&-2.05&$8.95\times 10^{-3}$&-2.16&$6.86\times 10^{-3}$&	1.90 	&	-1.41 	&	0.76 	\\
     ${}^{287}\mathrm{Mc}\to{}^{283}\mathrm{Nh}$&10.76&$0  ^+\to0  ^+$&0&-2.39&$4.04\times 10^{-3}$&-2.53&$2.96\times 10^{-3}$&	-1.22 	&	-3.55 	&	-1.02 	\\
     ${}^{288}\mathrm{Mc}\to{}^{284}\mathrm{Nh}$&10.65&$0  ^+\to0  ^+$&0&-2.99&$1.03\times 10^{-3}$&-3.11&$7.72\times 10^{-4}$&	-0.75 	&	-3.29 	&	-0.17 	\\
     ${}^{289}\mathrm{Mc}\to{}^{285}\mathrm{Nh}$&10.49&$0  ^+\to0  ^+$&0&-2.45&$3.56\times 10^{-3}$&-2.52&$3.02\times 10^{-3}$&	-0.39 	&	-2.87 	&	-0.35 	\\
     ${}^{290}\mathrm{Mc}\to{}^{286}\mathrm{Nh}$&10.41&$0  ^+\to0  ^+$&0&-2.97&$1.07\times 10^{-3}$&-3.17&$6.80\times 10^{-4}$&	-0.08 	&	-2.67 	&	0.50 	\\    
     ${}^{291}\mathrm{Lv}\to{}^{287}\mathrm{Fl}$&10.89&$0  ^+\to0  ^+$&0&-2.51&$3.10\times 10^{-3}$&-2.59&$2.60\times 10^{-3}$&	-1.59 	&	-3.64 	&	-1.06 	\\
     ${}^{293}\mathrm{Lv}\to{}^{289}\mathrm{Fl}$&10.68&$0  ^+\to0  ^+$&0&-2.57&$2.69\times 10^{-3}$&-2.58&$2.61\times 10^{-3}$&	-1.10 	&	-3.11 	&	-0.52 	\\
     ${}^{293}\mathrm{Ts}\to{}^{289}\mathrm{Mc}$&11.32&$0  ^+\to0  ^+$&0&-2.50&$3.20\times 10^{-3}$&-2.65&$2.22\times 10^{-3}$&	-1.60 	&	-4.42 	&	-1.77 	\\
     ${}^{294}\mathrm{Ts}\to{}^{290}\mathrm{Mc}$&11.18&$0  ^+\to0  ^+$&0&-3.09&$8.15\times 10^{-4}$&-3.21&$6.14\times 10^{-4}$&	-1.15 	&	-4.12 	&	-0.91 	\\
     \hline\hline
     \label{TBALE3}
     \end{tabular}
     \end{table*}

\begin{figure}[htbp]
  \centering
  \includegraphics[width=0.495\textwidth]{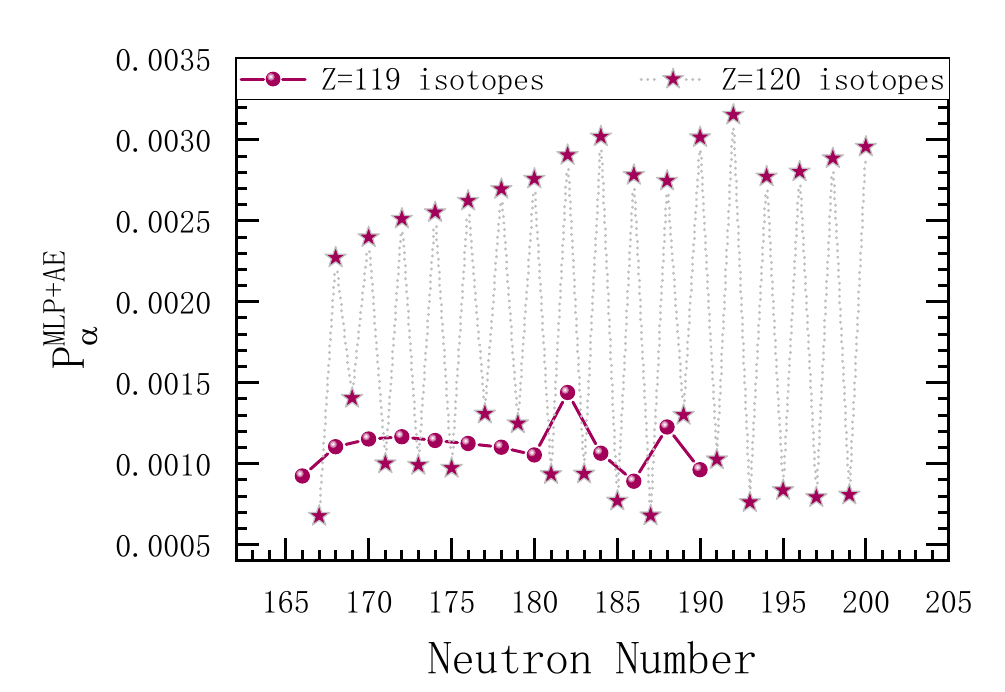}
  \caption{Predicted preformation factors for $Z$=119 and 120 isotope chains by MLP + AE method.}
  \label{Z = 119120}
\end{figure}

\begin{table*}[htbp]
\caption{Predictions of $\alpha$-decay half-lives of superheavy nuclei with $Z$=119 and 120. The decay energies are taken from Ref.~\cite{MaCPC}. The decay energies and half-lives are in the unit of `MeV' and `s', respectively.}
\begin{center}\begin{tabular}{ccccccccccccc}
  \hline\hline
$\alpha$ transition&~~~~$Q_\alpha$~~~~&~ ~$j_p^\pi\rightarrow j_d^\pi$~~&~~$l_\text{min}$&$\log_{10}P_\alpha^\text{MLP+AE}$&$\log_{10}T_{1/2}^\text{pred}$&$\alpha$ transition&~~~~$Q_\alpha$~~~~&~ ~$j_p^\pi\rightarrow j_d^\pi$~~&~~$l_\text{min}$&$\log_{10}P_\alpha^\text{MLP+AE}$&$\log_{10}T_{1/2}^\text{pred}$\\\hline
$^{285}119\rightarrow^{281}\text{Ts}$	&	14.07 	&	$1/2^{-}\rightarrow3/2^{-}$	&	2	&	-3.03 	&	-5.92 	&	$^{298}120\rightarrow^{294}\text{Og}$	&	13.09 	&	$0^{+}\rightarrow0^{+}$	&	0	&	-2.57 	&	-2.74 	\\
$^{287}119\rightarrow^{283}\text{Ts}$	&	13.26 	&	$1/2^{-}\rightarrow3/2^{-}$	&	2	&	-2.96 	&	-4.63 	&	$^{299}120\rightarrow^{295}\text{Og}$	&	13.10 	&	$1/2^{+}\rightarrow1/2^{+}$	&	0	&	-2.90 	&	-0.50 	\\
$^{289}119\rightarrow^{285}\text{Ts}$	&	12.95 	&	$1/2^{-}\rightarrow3/2^{-}$	&	2	&	-2.94 	&	-4.12 	&	$^{300}120\rightarrow^{296}\text{Og}$	&	13.05 	&	$0^{+}\rightarrow0^{+}$	&	0	&	-2.56 	&	-4.42 	\\
$^{291}119\rightarrow^{287}\text{Ts}$	&	12.76 	&	$1/2^{-}\rightarrow3/2^{-}$	&	2	&	-2.93 	&	-3.78 	&	$^{301}120\rightarrow^{297}\text{Og}$	&	12.90 	&	$3/2^{+}\rightarrow1/2^{+}$	&	2	&	-3.03 	&	-5.35 	\\
$^{293}119\rightarrow^{289}\text{Ts}$	&	12.70 	&	$1/2^{-}\rightarrow3/2^{-}$	&	2	&	-2.94 	&	-3.66 	&	$^{302}120\rightarrow^{298}\text{Og}$	&	12.81 	&	$0^{+}\rightarrow0^{+}$	&	0	&	-2.54 	&	0.19 	\\
$^{295}119\rightarrow^{291}\text{Ts}$	&	12.62 	&	$1/2^{-}\rightarrow3/2^{-}$	&	2	&	-2.95 	&	-3.57 	&	$^{303}120\rightarrow^{299}\text{Og}$	&	12.74 	&	$1/2^{+}\rightarrow3/2^{+}$	&	2	&	-3.03 	&	3.01 	\\
$^{297}119\rightarrow^{293}\text{Ts}$	&	12.55 	&	$1/2^{-}\rightarrow3/2^{-}$	&	2	&	-2.96 	&	-3.47 	&	$^{304}120\rightarrow^{300}\text{Og}$	&	12.66 	&	$0^{+}\rightarrow0^{+}$	&	0	&	-2.52 	&	-5.73 	\\
$^{299}119\rightarrow^{295}\text{Ts}$	&	12.60 	&	$1/2^{-}\rightarrow3/2^{-}$	&	2	&	-2.98 	&	-3.56 	&	$^{305}120\rightarrow^{301}\text{Og}$	&	13.45 	&	$3/2^{+}\rightarrow1/2^{+}$	&	2	&	-3.11 	&	-5.29 	\\
$^{301}119\rightarrow^{297}\text{Ts}$	&	12.39 	&	$3/2^{-}\rightarrow3/2^{-}$	&	0	&	-2.84 	&	-3.51 	&	$^{306}120\rightarrow^{302}\text{Og}$	&	13.29 	&	$0^{+}\rightarrow0^{+}$	&	0	&	-2.56 	&	-4.94 	\\
$^{303}119\rightarrow^{299}\text{Ts}$	&	12.23 	&	$3/2^{-}\rightarrow1/2^{-}$	&	2	&	-2.97 	&	-2.86 	&	$^{307}120\rightarrow^{303}\text{Og}$	&	13.25 	&	$1/2^{+}\rightarrow7/2^{+}$	&	4	&	-3.17 	&	-5.03 	\\
$^{305}119\rightarrow^{301}\text{Ts}$	&	12.87 	&	$3/2^{-}\rightarrow1/2^{-}$	&	2	&	-3.05 	&	-4.16 	&	$^{308}120\rightarrow^{304}\text{Og}$	&	13.07 	&	$0^{+}\rightarrow0^{+}$	&	0	&	-2.56 	&	-4.95 	\\
$^{307}119\rightarrow^{303}\text{Ts}$	&	12.60 	&	$1/2^{-}\rightarrow1/2^{-}$	&	0	&	-2.91 	&	-4.00 	&	$^{309}120\rightarrow^{305}\text{Og}$	&	12.12 	&	$5/2^{+}\rightarrow5/2^{+}$	&	0	&	-2.89 	&	-4.85 	\\
$^{309}119\rightarrow^{305}\text{Ts}$	&	11.58 	&	$5/2^{+}\rightarrow1/2^{-}$	&	3	&	-3.02 	&	-1.27 	&	$^{310}120\rightarrow^{306}\text{Og}$	&	11.61 	&	$0^{+}\rightarrow0^{+}$	&	0	&	-2.52 	&	-4.80 	\\
$^{287}120\rightarrow^{283}\text{Og}$	&	14.10 	&	$7/2^{+}\rightarrow1/2^{+}$	&	4	&	-3.17 	&	-5.12 	&	$^{311}120\rightarrow^{307}\text{Og}$	&	11.28 	&	$7/2^{+}\rightarrow5/2^{+}$	&	2	&	-2.99 	&	-4.39 	\\
$^{288}120\rightarrow^{284}\text{Og}$	&	13.72 	&	$0^{+}\rightarrow0^{+}$	&	0	&	-2.64 	&	-4.90 	&	$^{312}120\rightarrow^{308}\text{Og}$	&	10.79 	&	$0^{+}\rightarrow0^{+}$	&	0	&	-2.50 	&	-4.20 	\\
$^{289}120\rightarrow^{285}\text{Og}$	&	13.36 	&	$1/2^{+}\rightarrow1/2^{+}$	&	0	&	-2.85 	&	-4.51 	&	$^{313}120\rightarrow^{309}\text{Og}$	&	13.13 	&	$9/2^{+}\rightarrow7/2^{+}$	&	2	&	-3.12 	&	-5.38 	\\
$^{290}120\rightarrow^{286}\text{Og}$	&	13.44 	&	$0^{+}\rightarrow0^{+}$	&	0	&	-2.62 	&	-4.34 	&	$^{314}120\rightarrow^{310}\text{Og}$	&	11.81 	&	$0^{+}\rightarrow0^{+}$	&	0	&	-2.56 	&	-4.99 	\\
$^{291}120\rightarrow^{287}\text{Og}$	&	13.33 	&	$3/2^{+}\rightarrow1/2^{+}$	&	2	&	-3.00 	&	-4.29 	&	$^{315}120\rightarrow^{311}\text{Og}$	&	13.47 	&	$9/2^{+}\rightarrow9/2^{+}$	&	0	&	-3.08 	&	-1.97 	\\
$^{292}120\rightarrow^{288}\text{Og}$	&	13.22 	&	$0^{+}\rightarrow0^{+}$	&	0	&	-2.60 	&	-4.47 	&	$^{316}120\rightarrow^{312}\text{Og}$	&	11.31 	&	$0^{+}\rightarrow0^{+}$	&	0	&	-2.55 	&	0.09 	\\
$^{293}120\rightarrow^{289}\text{Og}$	&	13.22 	&	$5/2^{+}\rightarrow3/2^{+}$	&	2	&	-3.00 	&	-4.55 	&	$^{317}120\rightarrow^{313}\text{Og}$	&	11.10 	&	$5/2^{-}\rightarrow9/2^{+}$	&	3	&	-3.10 	&	-2.46 	\\
$^{294}120\rightarrow^{290}\text{Og}$	&	13.23 	&	$0^{+}\rightarrow0^{+}$	&	0	&	-2.59 	&	-3.86 	&	$^{318}120\rightarrow^{314}\text{Og}$	&	10.65 	&	$0^{+}\rightarrow0^{+}$	&	0	&	-2.54 	&	-1.30 	\\
$^{295}120\rightarrow^{291}\text{Og}$	&	13.17 	&	$1/2^{+}\rightarrow5/2^{+}$	&	2	&	-3.01 	&	-3.56 	&	$^{319}120\rightarrow^{315}\text{Og}$	&	10.09 	&	$1/2^{+}\rightarrow5/2^{-}$	&	3	&	-3.09 	&	0.43 	\\
$^{296}120\rightarrow^{292}\text{Og}$	&	13.16 	&	$0^{+}\rightarrow0^{+}$	&	0	&	-2.58 	&	-4.90 	&	$^{320}120\rightarrow^{316}\text{Og}$	&	10.06 	&	$0^{+}\rightarrow0^{+}$	&	0	&	-2.53 	&	2.11 	\\
$^{297}120\rightarrow^{293}\text{Og}$	&	13.06 	&	$1/2^{+}\rightarrow1/2^{+}$	&	0	&	-2.88 	&	-4.04 	&--&--&--&--&--&--\\\hline\hline
\end{tabular}\label{pre}
\end{center}
\end{table*}

\subsection{Prediction of superheavy nuclei with $Z = 119$ and $Z = 120$}\label{Summary}
Accurate prediction of the $\alpha$-decay half-life is crucial for the synthesis and structural research of these elements. The improved MLP model has been well trained and demonstrates a reliable predictive capability for preformation factors. Consequently, it is used to calculate the $\alpha$ formation amplitude for nuclei within the superheavy region. In the final selected model, the test set value closest to the mean is selected to predict the preformation factors of superheavy nuclei with Z = 119 and 120. The decay energies $Q_{\alpha}$ are taken from Ref.~\cite{MaCPC}. The predictions are restricted to even-even nuclei and odd-$A$ nuclei at $Z$=119 and 120 due to the challenges associated with calculating the minimum angular momentum transferred to the \(\alpha\)-particle of odd-odd nuclei. The prediction results are shown in Fig. \ref{Z = 119120} and exhibit a notable odd-even staggering effect. For example, when the neutron number is 182, the preformation factor $P_\alpha^\text{MLP+AE}$ of the nucleus ($Z = 120, N = 182$) is larger than that of the nucleus ($Z = 119, N = 182$), which is approximately reduced from 0.0029 to 0.0014, a decrease of approximately 52$\%$. For the $Z = 120$ isotope chain, $P_\alpha^\text{MLP+AE}$ of the nucleus of $N = 182$ is larger than that of $N = 183$ ($P_\alpha^\text{MLP+AE}$=0.00094), indicating that unpaired proton and neutron will inhibit the formation of $\alpha$-particle during the decay process. In addition, the overall preformation factor of the $Z = $120 isotope chain is greater than that of $Z = $119. The predicted values of the half-lives of the $Z = $119 and 120 isotope chains are listed in Table \ref{pre} to provide a reference for experimental synthesis of new nuclides.


\section{SUMMARY}\label{SUMMARY}
We incorporate an autoencoder into the traditional multi-layer perceptron (MLP) neural network. By adjusting the free parameter $\beta$, we can lock in the optimal parameters for different types of even-even, odd-$A$ and odd-odd nuclei. Our study found that the preformation factor calculated by the MLP + AE method can give an accuracy comparable to the empirical formula. At the same time, the MLP + AE method can effectively learn the relationship between the preformation factor and the decay energy, that is, the linear relationship between $\text{log}_{10}P_\alpha^\text{MLP+AE}$ and $Q_\alpha^{-1/2}$. This again supports the notion that the Geiger-Nuttall law can be used to deal with preformation factors. We also studied changes in preformation factors in isotope chains. The preformation factor is found to show different trends on both sides of the magic number, which also reflects that the shell effect directly affects the preformation of the $\alpha$ particle. Furthermore, as the number of protons increases, the preformation factor exhibits odd-even staggering. A similar situation also occurs on isotone chains. This means that unpaired nuclei inhibit the preformation of $\alpha$ particles on the surface of the parent nucleus. This also reflects that preformation factors can provide important information for the study of nuclear structure. In addition, we used the MLP + AE neural network to calculate the preformation factors of 41 nuclei in the AME2020 mass table and find that the preformation factors calculated by MLP + AE were equivalent to the results calculated by the empirical formula. The MLP + AE method has been proven to make reliable predictions of preformation factors. Finally, the MLP + AE method was used to calculate the preformation factors of the two isotope chains $Z = $119 and 120, and their half-lives were further predicted, providing a reliable theoretical reference for the experimental synthesis of $Z>$118 elements.

\begin{acknowledgments}

  Nana Ma is supported by National Key Research and Development (R\&D) Program under Grant No. 2021YFA1601500 and the National Natural Science Foundation of China under Grant No. 12105128. Jungang Deng is supported by the National Natural Science Foundation of China under Grant No. 12405138, Hubei Provincial Natural Science Foundation of China (Grant No. 2023AFB035), and Natural Science Research Project of Yichang City (Grant No. A23-2-028).

\end{acknowledgments}

\end{document}